\documentclass[11pt]{amsart}
\DeclareMathAlphabet{\EuFrak}{U}{euf}{m}{n}
\DeclareMathAlphabet{\EuScript}{U}{eus}{m}{n}
\usepackage{amsmath}
\usepackage{amsfonts,amssymb,latexsym}
\usepackage{mathrsfs}
\usepackage{amscd}
\usepackage{amsthm}
\newtheorem{theorem}{\rmfamily\bfseries{Theorem}}[section] 
\newtheorem{corollary}[theorem]{\rmfamily\bfseries{Corollary}} 

\newtheorem{lemma}[theorem]{\rmfamily\bfseries{Lemma}} 
\newtheorem{proposition}[theorem]{\rmfamily\bfseries{Proposition}}

\newtheorem{definition}[theorem]{\rmfamily\bfseries{Definition}}

\theoremstyle{remark}

\newtheorem{remark}{Remark}

\usepackage{graphics}
\numberwithin{equation}{section}
\def\Bbb{\mathbb}
\newcommand{\mona}{\mbox{\LARGE\itshape a}}
\def\logo{\raisebox{-10.5\p@}{\hb@xt@85\p@{\includegraphics{gft.eps}\hfil}}}

\def\un{1\kern-3pt \rm I}

\newcommand{\oN}{{\mathbb N}}
\newcommand{\oR}{{\mathbb R}}

\newcommand{\oC}{{\mathbb C}}

\newcommand{\supp}{\mathrm{supp}}
   
\begin{document}

\title[NCQFT in Terms of Tempered Ultrahyperfunctions]
      {\sl{Ultrahyperfunctional Approach to \\ Non-Commutative Quantum Field Theory}}

\author{Daniel H.T. Franco}
\address{Centro de Estudos de F\'\i sica Te\'orica, Setor de F\'\i sica--Matem\'atica\\
         Rua Rio Grande do Norte 1053/302, Funcion\'arios \\
         Belo Horizonte, Minas Gerais, Brasil, CEP:30130-131.}
\email{dhtfranco@gmail.com}

\author{Jos\'e A. Louren\c co}
\address{Universidade Federal do Esp\'\i rito Santo\\
         Departamento de F\'\i sica\\  
         Campus Universit\'ario de Go\-i\-a\-be\-iras, Vit\'oria, ES, Brasil, CEP:29060-900.}
\email{quantumlourenco@gmail.com}

\author{Luiz H. Renoldi}
\address{Centro de Estudos de F\'\i sica Te\'orica, Setor de F\'\i sica--Matem\'atica\\
         Rua Rio Grande do Norte 1053/302, Funcion\'arios \\
         Belo Horizonte, Minas Gerais, Brasil, CEP:30130-131.}
\email{lhrenoldi@yahoo.com.br}

\keywords{Non-commutative theory, axiomatic field theory, tempered ultrahyperfunctions.}
\subjclass{46F15, 46F20, 81T05}
\date{\today}
\thanks{J.A. Louren\c co is supported by the Brazilian agency CNPq}

\begin{abstract}
In the present paper, we intent to enlarge the axiomatic framework of non-commutative
quantum field theories (QFT). We consider QFT on non-commutative spacetimes in terms
of the tempered ultrahyperfunctions of Sebasti\~ao e Silva corresponding to a convex
cone, within the framework formulated by Wightman. Tempered ultrahyperfunctions are
representable by means of holomorphic functions. As is well known there are certain
advantages to be gained from the representation of distributions in terms of holomorphic
functions. In particular, for non-commutative theories the Wightman functions involving
the $\star$-product, ${\mathfrak W}^\star_m$, have the same form as the standard
form ${\mathfrak W}_m$. We conjecture that the functions
${\mathfrak W}^\star_m$ satisfy a set of properties which actually will characterize
a non-commutative QFT in terms of tempered ultrahyperfunctions.
In order to support this conjecture, we prove for this setting the validity of some
important theorems, of which the CPT theorem and the theorem on the Spin-Statistics
connection are the best known. We assume the validity of these theorems for non-commutative
QFT in the case of spatial non-commutativity only.
\end{abstract}

\maketitle

\centerline{\sf Dedicated to Prof. Olivier Piguet on the occasion of his 65th birthday.}

\section{Introduction}
In recent years, many novel questions have emerged in theoretical physics,
particularly in non-commutative quantum field theories (NCQFT), for which a
considerable effort has been made in order to clarify structural aspects from
an axiomatic standpoint~\cite{AGVM}-\cite{Solo5}.
Axiomatic Quantum Field Theory is the program, originally conceived by G\r{a}rding
and Wightman~\cite{SW}-\cite{Haag}, that aims to study of unified form the fundamental
postulates, and their consequences, of the two pillars apparently opposite of the modern
physics: the Relativity Theory and the Quantum Mechanics. The standard formulation of the
axioms of quantum field theories is best expressed by the so-called Wightman axioms, which
can be summarized as follows: (I) {\em Quantum mechanical postulates}. The states are
described by vectors of a Hilbert space $\mathscr H$. In $\mathscr H$, there exists
a unitary representation of the Poincar\'e group, whose translation group admits
the closed forward light cone $\overline{V}_+=\bigl\{p_\mu \in \oR^4 \mid p^2 \geq 0,
p^0 \geq 0\bigr\}$ as its spectrum. There is a unique vaccum state $|\Omega_o\rangle$ in
$\mathscr H$, which is the unique state invariant by translations (this implies in the
uniqueness of the vacuum). (II) {\em Special relativity postulates}. The fields transform
covariantly under Poincar\'e transformations. The microcausality condition imposes
that the fields either commute or anti-commute at spacelike separated points
$\bigl[\Phi(x),\Phi(x^\prime)\bigr]_{\pm}=0$ for $(x-x^\prime)^2<0$.
(III) {\em Technical postulate}. The assumption of a character of distribution takes
essential place among the basis postulates of quantum field theory. In a mathematical
language, there are some reasons to consider the fields as {\it tempered}
distributions~\cite{SW}-\cite{Haag}. This choice is connected with a definition of
{\em local} properties of distributions. It turn out that all these
postulates can be fully reexpressed in terms of an infinite set of tempered
distributions, called Wightman distributions (or correlation functions of the theory).

By a variety of reasons, the Wightman framework of local QFT turned out to be too narrow
for theoretical physicists, who are interested in handling situations involving in
particular NCQFT. One of the reasons is that the commutation relations for the
{\em non-commutative coordinates} $[x_\mu,x_\nu]=i \theta_{\mu\nu}$ break down the Lorentz
group $SO(1,3)$ to a {\em residual} symmetry $SO(1,1)\times SO(2)$. This happens because the
deformation parameter $\theta_{\mu\nu}$ is assumed to be a {\em constant} antisymmetric matrix
of lenght dimension two. Although an axiomatic formulation has been proposed based in
the {\em residual} symmetry $SO(1,1)\times SO(2)$~\cite{AGVM}-\cite{Solo5}, a serious
inconvenient arises of this analysis: the subgroup $SO(1,1)\times SO(2)$ does not allow
that particles be classified according to the 4-dimensional Wigner particle
concept~\cite{Bert}-\cite{ChaTu}.

Another reason why the framework of local QFT turned out to be too narrow, it
is that NCQFT are {\em nonlocal}. This can have implications on highly physical
properties. For example, in the formulation of general
properties of a field theory the localization plays a fundamental role in the concrete
realization of the locality of field operators in coordinate space and spectral condition
in energy-momentum space, which are achieved through the {\em localization of test
functions} -- the fields are considered tempered functionals on the Schwartz's test function
space, the space of rapidly decreasing $C^\infty$-functions. However, the nonlocal character
of the interactions in NCQFT seems to indicate that fields are not tempered.
In fact, as it was emphasized in~\cite{AGVM}, the existence of hard infrared
singularities in the non-planar sector of the theory, induced by uncancelled quadratic
ultraviolet divergences, can destroy the {\em tempered} nature of the Wightman functions.
Besides, the commutation relations $[x_\mu,x_\nu]=i \theta_{\mu\nu}$
also imply uncertainty relations for spacetime coordinates
$\Delta x_\mu \Delta x_\nu \sim \bigl|\theta_{\mu\nu}\bigr|$, indicating
that the notion of spacetime point loses its meaning. Spacetime points are replaced by
cells of area of size $\bigl|\theta_{\mu\nu}\bigr|$. This observation has led physicists
to suggest the existence of a finite lower limit to the possible resolution of distance.
Instead, the nonlocal structure of NCQFT manifests itself in the {\em delocalization} of the
interaction regions, which spread over a spacetime domain whose size is determined by
the existence of a {\em minimum length} $\ell_\theta$ related to the scale of nonlocality
$\ell_\theta \sim \sqrt{\theta}$~\cite{Szabo}. Among other things, the existence of this
{\em minimum length} renders impossible the preservation of the local commutativity
condition, so it is unclear why we should even consider the microcausal condition based
on local fields as in~\cite{AGVM,Chai1,Green}. 

These are some very important evidences to expect that the traditional Wightman
axioms must be somewhat modified within the context of NCQFT~\cite{Comment}.
From our point of view, the spacetime non-commutativity can be accommodated
simply by choosing a space of generalized functions different from the
usual space of Schwartz's tempered distributions. As a matter
of fact, in a fundamental formulation of QFT, the mathematical problem can be seen
as a problem of the choice of the {\it right} class of generalized functions
which is appropriate for the representation of quantum fields. Thus, the class of
generalized functions which one should use in the formulation
of NCQFT remains an open problem still to be fully understood.

Some attempts have been made to extend the framework fomulated by Wightman for NCQFT,
so as to include a wider class of fields~\cite{DanCaio,Solo5}. It has been suggested
that NCQFT must should be formulated in terms of generalized functions over the space of
analytic test functions ${\sf S}^0$~\cite{Luc1}-\cite{Solo4}, exploring some ideas by
Soloviev to nonlocal quantum fields~\cite{Solo1}-\cite{Solo4}.\footnote{More recently,
Chaichian {\it et al}~\cite{CMTV} have obtained a result that the appropriate space of test
functions in the Wightman approach to non-commutative quantum field theory is one of the
Gel'fand-Shilov spaces ${\sf S}^\beta$, with $\beta < 1/2$~\cite{GelShi}. The authors of
Refs.~\cite{DanCaio,Solo5} assume $\beta=0$ in order to emphasize that this is the smallest
space among the Gel'fand-Shilov spaces $\sf S^\beta$ traditionally adopted in
nonlocal quantum field theory, as indicated from non-commutative quantum field theory.}
In this case, the fields are so singulars that, of course, one of the conceptual problems
we are faced is find an adequate generalization of the causality condition. Soloviev has
suggested to replace the ordinary causality condition by an asymptotic causality condition.
Despite its apparent weakness, the asymptotic causality condition in the sense of Soloviev
yet one allows us to show the validity of the CPT theorem and the Spin-Statistic connection
for NCQFT~\cite{DanCaio}. And more, the existence of a Borchers class for a non-commutative
field is shown~\cite{Daniel}. On the other hand, recently, different definitions of
perturbative theory to NCQFT~\cite{Bahns,Manfred} seem to point out that the nonlocal interactions
in NCQFT improve the UV behavior of theory. It is therefore reasonable to consider another space
of test functions where the fields are not highly singulars as adopted in~\cite{DanCaio,Solo5}.

In this paper, we present an alternative approach. Because NCQFT suggest the existence of
a minimum length $\ell_\theta$, we will assume as space of test functions for NCQFT the space
$\mathfrak H$ of rapidly decreasing entire functions in any horizontal strip. The elements of
the dual space of the space $\mathfrak H$ are so-called tempered
ultrahyperfunctions~\cite{Tiao1}-\cite{Daniel1} and have the advantage of being representable
by means of holomorphic functions. Tempered ultrahyperfunctions generalize the notion of
hyperfunctions on $\oR^n$ but {\em can not} be localized as hyperfunctions. Because of this,
NCQFT of this sort will be called {\em quasilocal}, namely, the fields
are localizable only in regions greater than the scale of nonlocality $\ell_\theta$.
We shall walk along the general lines proposed recently by Br\"uning-Nagamachi~\cite{BruNa1}.
They have conjectured that tempered ultrahyperfunctions, {\em i.e.}, those ultrahyperfunctions
which admit the Fourier transform as an isomorphism of topological vector spaces, are well
adapted for their use in quantum field theory with a fundamental length.
In particular, we shall consider tempered ultrahyperfunctions in a setting which includes
the results of~\cite{Tiao1,Tiao2,Hasumi} as special cases, by considering
functions analytic in tubular radial domains~\cite{Carmi2,DanHenri,Daniel1}.
We shall denote the NCQFT in terms of tempered ultrahyperfunctions by UHFNCQFT for
brevity, hereafter.

The presentation of the paper is organized as follows. In Section \ref{SecMot}, for the
convenience of the reader, we present the reasons why tempered ultrahyperfunctions are
well adapted for their use in NCQFT, going through a simple example taken from Ref.~\cite{BruNa1}.
Section \ref{SecUltra} contains an exposition of the theory of tempered ultrahyperfuntions,
where we include and prove some results which are important in applications to quantum field theory.
Section \ref{SecAxioms} is devoted to the formulation of the axioms for UHFNCQFT in terms
of the Wightman functionals. How the properties of the Wightman functionals
change when we pass to the test function space which are entire analytic functions
of rapid decrease in any horizontal strip is considered. In Section \ref{SecTheorems}, we derive for
our UHFNCQFT the validity of some important theorems, obtained previously for essentially
nonlocalizable fields~\cite{DanCaio,Daniel,Solo5}. These include
the existence of CPT symmetry and the connection between Spin and Statistics
for UHFNCQFT. Throughout the paper we assume only the case of space-space non-commutativity,
{\em i.e.}, $\theta_{0i}=0$, with $i=1,2,3$. It is well known that if there is space-time
non-commutativity, the resulting theory violates the causality and unitarity~\cite{SST,Gomis}.
For most our purposes, we consider for simplicity a theory with only one basic field, a
neutral scalar field. Section \ref{SecFinal} is reserved for our concluding remarks.

\section{Motivation}
\label{SecMot}
For the sake of completeness in the exposition, we recall the example which has motivated
Br\"uning-Nagamachi~\cite{BruNa1} to conjecture that tempered ultrahyperfunctions are suitable
in order to treat quantum field theories with a minimum length. Consider the Dirac delta
measure $\delta(x+a)$, which when applied to a continuous function $f(x)$ produces the
value $f(-a)$
\[
\int \delta(x+a)f(x)\,\,dx=f(-a)\,\,.
\] 
By using a generalization of the Cauchy's integral formula, we define $\delta(x+a)$
applied to a holomorphic function $f(z)$ on an open set $\Omega \subset \oC$.
Assuming that $0 \in \Omega$ and letting $\gamma=\partial \Omega$ denote the
boundary of $\Omega$, we have
\begin{equation}
\frac{1}{2 \pi i} \oint_\gamma \frac{f(z)}{z+a}\,\,dz=f(-a)\,\,,
\quad{\mbox{for}}\,\,z \in \Omega\,\,.
\label{Cauchy}
\end{equation}
Define ${\mathfrak H}(T(-\ell,\ell))$ as being the space of all
holomorphic functions $f(z)$ on $T(-\ell,\ell)=\oR^n+i(-\ell,\ell) \subset \oC$.
In this case, from (\ref{Cauchy}), for $f(z) \in {\mathfrak H}(T(-\ell,\ell))$
and $|a|<\ell$, $f(-a)$ can be given by the Taylor's series of center in zero
\[
f(-a)=\sum_{n=0}^\infty \frac{(-a)^n}{n!}f^{(n)}(0)\,\,.
\]
This series possesses the functional representation
\begin{align*}
F(f)&=\int \Bigl[\sum_{n=0}^\infty \frac{a^n}{n!}\delta^{(n)}(x)\Bigr]f(x)\,\,dx
=\sum_{n=0}^\infty \frac{(-a)^n}{n!}f^{(n)}(0)\\[3mm]
&=f(-a)=\int \delta(x+a)f(x)\,\,dx\,\,.
\end{align*}
Thus, as an equation for functionals defined on the function space
${\mathfrak H}(T(-\ell,\ell))$, we have the identification
\begin{align*}
\sum_{n=0}^\infty \frac{a^n}{n!}\delta^{(n)}(x)=\delta(x+a)\,\,,
\end{align*}
in the distributional sense. In other words, the sequence of generalized
functions
\[
S_N=\sum_{n=0}^N \frac{a^n}{n!}\delta^{(n)}(x)\,\,,
\]
with support $\{0\}$ weakly converges to the generalized function $\delta(x+a)$
with support $\{-a\}$, as $N \rightarrow \infty$. However, if $|a|>\ell$, this
sequence does not converge in the dual space of ${\mathfrak H}(T(-\ell,\ell))$.

The motivation for suggesting that tempered ultrahyperfunctions are well adapted
for their use in quantum field theory with a fundamental length lies in
the following fact: the non-local structure of the functional $F$ is represented
by a dislocation of the support from $\{0\}$ to $\{-a\}$. According to
Br\"uning-Nagamachi~\cite{BruNa1}, this means that, if $|a|<\ell$, then the elements
in the dual space of ${\mathfrak H}(T(-\ell,\ell))$ do not distinguish between the
points $\{0\}$ to $\{-a\}$, but if $|a|>\ell$ the elements in
${\mathfrak H}^\prime(T(-\ell,\ell))$ can distinguish between the points
$\{0\}$ to $\{-a\}$. Since $|a|<\ell$ is arbitrary, one can say that the elements
in ${\mathfrak H}^\prime(T(-\ell,\ell))$ distinguish points only in spacetime
regions large in comparison with $\ell$. This is the reason why we discuss here a
mathematically more satisfactory approach for NCQFT. The tempered ultrahyperfunctions
have this property.       
 
\begin{remark}
Such an example was already considered in 1958 by G\"uttinger~\cite{Gut} in order to treat
certain exactly soluble models which would correspond to field theories with
non-renormalizable interactions.
\end{remark} 

\section{Tempered Ultrahyperfunctions}
\label{SecUltra}
The interest in tempered ultrahyperfunctions arose simultaneously with the
growing interest in various classes of analytic functionals and various attempts
to develop a theory of such functionals which would be analogous to the
Schwartz theory of distributions. Tempered ultrahyperfunctions were first introduced
in papers of Sebasti\~ao e Silva~\cite{Tiao1,Tiao2} and Hasumi~\cite{Hasumi} as
the strong dual of the space of test functions ${\mathfrak H}$ of rapidly
decreasing entire functions in any horizontal strip. As a matter of fact, these
objects are equi\-va\-lence classes of holomorphic functions defined by a certain
space of functions which are analytic in the $2^n$ octants in $\oC^n$ and represent
a natural generalization of the notion of hyperfunctions on $\oR^n$, but are
{\it non-localizable}. In this section, we recall some basic pro\-per\-ties of the
tempered ultrahyperfunction space which are the most important in
applications to quantum field theory.

To begin with, we shall define our notation. We will use the standard multi-index notation.
Let $\oR^n$ (resp. $\oC^n$) be the real (resp. complex) $n$-space whose generic points
are denoted by $x=(x_1,\ldots,x_n)$ (resp. $z=(z_1,\ldots,z_n)$), such that
$x+y=(x_1+y_1,\ldots,x_n+y_n)$, $\lambda x=(\lambda x_1,\ldots,\lambda x_n)$,
$x \geq 0$ means $x_1 \geq 0,\ldots,x_n \geq 0$, $\langle x,y \rangle=x_1y_1+\cdots+x_ny_n$
and $|x|=|x_1|+\cdots+|x_n|$. Moreover, we define
$\alpha=(\alpha_1,\ldots,\alpha_n) \in \oN^n_o$, where $\oN_o$ is the set
of non-negative integers, such that the length of $\alpha$ is the corresponding
$\ell^1$-norm $|\alpha|=\alpha_1+\cdots +\alpha_n$, $\alpha+\beta$ denotes
$(\alpha_1+\beta_1,\ldots,\alpha_n+\beta_n)$, $\alpha \geq \beta$ means
$(\alpha_1 \geq \beta_1,\ldots,\alpha_n \geq \beta_n)$, $\alpha!=
\alpha_1! \cdots \alpha_n!$, $x^\alpha=x_1^{\alpha_1}\ldots x_n^{\alpha_n}$,
and
\[
D^\alpha \varphi(x)=\frac{\partial^{|\alpha|}\varphi(x_1,\ldots,x_n)}
{\partial x_1^{\alpha_1}\partial x_2^{\alpha_1}\ldots\partial x_n^{\alpha_n}}\,\,.
\]
Let $\Omega$ be a set in $\oR^n$. Then we denote by $\Omega^\circ$ the interior
of $\Omega$ and by $\overline{\Omega}$ the closure of $\Omega$. For $r > 0$, we
denote by $B(x_o;r)=\bigl\{x \in \oR^n \mid |x-x_o| < r\bigr\}$ a open ball
and by $B[x_o;r]=\bigl\{x \in \oR^n \mid |x-x_o| \leq r\bigr\}$ a closed ball,
with center at point $x_o$ and of radius $r=(r_1,\ldots,r_n)$, respectively.

We consider two $n$-dimensional spaces -- $x$-space and $\xi$-space -- with the
Fourier transform defined
\[
\widehat{f}(\xi)={\mathscr F}[f(x)](\xi)=
\int_{\oR^n} f(x)e^{i \langle \xi,x \rangle} d^nx\,\,,
\]
while the Fourier inversion formula is
\[
f(x)={\mathscr F}^{-1}[\widehat{f}(\xi)](x)= \frac{1}{(2\pi)^n}
\int_{\oR^n} \widehat{f}(\xi)e^{-i \langle \xi,x \rangle} d^n\xi\,\,.
\]
The variable $\xi$ will always be taken real while $x$ will also be
complexified -- when it is complex, it will be noted $z=x+iy$. The
above formulas, in which we employ the symbolic ``function notation,''
are to be understood in the sense of distribution theory.

\,\,\,We shall consider the function
\[
h_{K}(\xi)=\sup_{x \in K} \bigl| \langle \xi,x \rangle \bigr|\,\,,\quad \xi \in \oR^n\,\,,
\]
the indicator of $K$, where $K$ is a compact set in $\oR^n$. $h_{K}(\xi) < \infty$
for every $\xi \in \oR^n$ since $K$ is bounded. For sets
$K=\bigl[-k,k\bigr]^n$, $0 < k < \infty$, the indicator function $h_{K}(\xi)$ can
be easily determined:
\[
h_{K}(\xi)=\sup_{x \in K} \bigl| \langle \xi,x \rangle \bigr|=
k|\xi|\,\,,\quad \xi \in \oR^n\,\,,\quad |\xi|=\sum_{i=1}^n|\xi_i|\,\,.
\]
Let $K$ be a convex compact subset of $\oR^n$,
then $H_b(\oR^n;K)$ ($b$ stands for bounded) defines the space of all
functions in $C^\infty(\oR^n)$ such that $e^{h_K(\xi)}D^\alpha f(\xi)$
is bounded in $\oR^n$ for any multi-index $\alpha$. One defines in
$H_b(\oR^n;K)$ seminorms
\begin{equation}
\|\varphi\|_{K,N}=\sup_{{\substack{\xi \in \oR^n \\ \alpha \leq N}}}
\bigl\{e^{h_K(\xi)}|D^\alpha f(\xi)|\bigr\} < \infty\,\,,
\quad N=0,1,2,\ldots\,\,.
\label{snorma2}
\end{equation}

Now, let $T(\Omega)=\oR^n+i\Omega \subset \oC^n$ be
the tubular set of all points $z$, such that $y_i={\text{Im}}\,z_i$ belongs
to the domain $\Omega$, {\em i.e.}, $\Omega$ is a connected open set in $\oR^n$
called the basis of the tube $T(\Omega)$. Let $K$ be a convex compact
subset of $\oR^n$, then ${\mathfrak H}_b(T(K))$ defines the space
of all $C^\infty$ functions $\varphi$ on $\oR^n$ which can be extended to
$\oC^n$ to be holomorphic functions in the interior $T(K^\circ)$ of $T(K)$ such
that the estimate
\begin{equation}
|\varphi(z)| \leq {\boldsymbol{\sf C}} (1+|z|)^{-N}
\label{est}
\end{equation}
is valid for some constant ${\boldsymbol{\sf C}}={\boldsymbol{\sf C}}_{K,N}(\varphi)$.
The best possible constants in (\ref{est}) are given by a family of seminorms in
${\mathfrak H}_b(T(K))$
\begin{equation}
\|\varphi\|_{T(K),N}=\sup_{z \in T(K)}
\bigl\{(1+|z|)^{N}|\varphi(z)|\bigr\} < \infty\,\,,
\quad N=0,1,2,\ldots\,\,.
\label{snorma1}
\end{equation}

Next, we consider a set of results which will characterize the spaces introduced above.

\begin{lemma}
If $K_i \subset K_{i+1}$ are two convex compact sets, then the following
canonical injections holds: (i) ${\mathfrak H}_b(T(K_{i+1}))
\hookrightarrow {\mathfrak H}_b(T(K_i))$, (ii) $H_b(\oR^n;K_{i+1})
\hookrightarrow H_b(\oR^n;K_i)$.
\label{injec1}
\end{lemma}

\begin{proof}
We prove the first item. If $K_i \subset K_{i+1}$ and
$\varphi \in {\mathfrak H}_b(T(K_{i+1}))$, then
$\varphi \in {\mathfrak H}_b(T(K_i))$. By taking the restriction of
$\varphi \in {\mathfrak H}_b(T(K_{i+1}))$ to $T(K_i)$, it follows that
\[
\sup_{z \in T(K_{i+1})}\bigl\{(1+|z|)^{j}|\varphi(z)|\bigr\}=
\sup_{z \in T(K_{i})}\bigl\{(1+|z|)^{j}|\varphi(z)|\bigr\}\,\,.
\]
Therefore, the topology induced by ${\mathfrak H}_b(T(K_{i+1}))$ on
${\mathfrak H}_b(T(K_i))$ is identical with the topology
of $\varphi \in {\mathfrak H}_b(T(K_i))$. The proof of second statement
is similar, taking into account the seminorm (\ref{snorma2}).
\end{proof}

Let $O$ be a convex open set of $\oR^n$. To define the topologies
of $H(\oR^n;O)$ and ${\mathfrak H}(T(O))$ it suffices to let $K$ range
over an increasing sequence of convex compact subsets $K_1,K_2,\ldots$
contained in $O$ such that for each $i=1,2,\ldots$,
$K_i \subset K_{i+1}^\circ$ and ${O}=\bigcup_{i=1}^\infty K_i$.
Then the spaces $H(\oR^n;O)$ and ${\mathfrak H}(T(O))$ are the projective
limits of the spaces $H_b(\oR^n;K)$ and ${\mathfrak H}_b(T(K))$, respectively,
{\em i.e.}, we have that
\begin{equation}
H(\oR^n;O)=\underset{K \subset {O}}{\lim {\rm proj}}\,\,
H_b(\oR^n;K)\,\,,
\label{limproj2}
\end{equation}
and
\begin{equation}
{\mathfrak H}(T({O}))=\underset{K \subset {O}}{\lim {\rm proj}}\,\,
{\mathfrak H}_b(T(K))\,\,,
\label{limproj1}
\end{equation}
where the projective limit is taken following the restriction mappings according
to the Lemma \ref{injec1}.

\begin{remark}
Any $C^\infty$ function of exponential growth is a multiplier in $H(\oR^n;O)$,
while that any $C^\infty$ function which can be extended to be an entire function
of polynomial growth is a multiplier in ${\mathfrak H}(T(O))$. Besides, the
space $H(\oR^n;O)$ is continuosly embedded into Schwartz space ${\mathscr S}(\oR^n)$,
and elements of ${\mathscr S}(\oR^n)$ are also multipliers for the space
$H(\oR^n;O)$~\cite{Hasumi}.
\end{remark}

\begin{lemma}
The spaces ${\mathfrak H}(T(O))$ and $H(\oR^n;O)$ are Hausdorff locally convex spaces.
\end{lemma}

\begin{proof}
First, we prove that ${\mathfrak H}(T(O))$ is a Hausdorff locally convex space. 
Let $\{K_i\}_{i=1,2,\ldots}$ be the usual increasing sequence of compact subsets of
$O$, whose union is $O$, and such that with $K_i$ is the closure of its interior,
$K_{i+1}^\circ$; for all $i$, $K_i \subset K_{i+1}^\circ$. We shall prove that each
element of the base for neighborhoods of 0 generated by the open balls
\[
{\mathfrak B}_{i,n}(0)=\Bigl\{\varphi \in {\mathfrak H}(T(K_i)) \mid
\|\varphi\|_{T(K_i),j}=\sup_{z \in T(K_i)} \Bigl[(1+|z|)^{j}|\varphi(z)|\Bigr]
< n^{-1}, n \in \oN\Bigr\}\,\,,
\]
contains at least one convex neighborhood of 0. For this, it is sufficient to show
that there exist natural numbers $\ell,n^\prime$ such that
${\mathfrak B}_{\ell,n^\prime}(0) \subset {\mathfrak B}_{i,n}(0)$. In fact, one
can always choose $\ell$ such that $K_\ell \subset K_i$. Then, $\|\varphi\|_{T(K_\ell),j}
\leq \|\varphi\|_{T(K_i),j}$ if $n < n^\prime$ and $\ell \leq i$. Now, consider
$\|\lambda \varphi_1 + (1-\lambda)\varphi_2\|_{T(O),j}$, with $0 \leq \lambda \leq 1$
and $\varphi_1, \varphi_2 \in {\mathfrak B}_{\ell,n^\prime}(0)$. But,
\begin{align*}
\|\lambda \varphi_1 + (1-\lambda)\varphi_2\|_{T(O),j} &\leq
\|\lambda \varphi_1\|_{T(O),j} + \|(1-\lambda)\varphi_2\|_{T(O),j} \\[3mm]
&\leq \lambda \|\varphi_1\|_{T(O),j} + (1-\lambda)\|\varphi_2\|_{T(O),j} \\[3mm]
& < \lambda n^{-1} + (1-\lambda) n^{-1} = n^{-1}\,\,.
\end{align*}
Hence, $\lambda \varphi_1 + (1-\lambda)\varphi_2 \in {\mathfrak B}_{\ell,n^\prime}(0)$.
This proves that ${\mathfrak H}(T(O))$ is locally convex.
Now, let $\varphi_1, \varphi_2, \psi \in {\mathfrak H}(T(O))$.
Consider that for the pair of distinct functions $\varphi_1, \varphi_2$,
$\|\varphi_1-\varphi_2\|_{T(O),j}=\varepsilon > 0$. Let $\boldsymbol{\phi}(\varphi_i)=
{\mathfrak B}_{\varepsilon/3}(\varphi_i)=\bigl\{\psi \in {\mathfrak H}(T(O)) \mid
\|\varphi_i-\psi\|_{T(O),j} < \varepsilon/3, i=1,2\bigr\}$. For if
$\psi \in \boldsymbol{\phi}(\varphi_1) \cap \boldsymbol{\phi}(\varphi_2)$,
we have $\|\varphi_1-\psi\|_{T(O),j}< \varepsilon/3$ and $\|\varphi_2-\psi\|_{T(O),j}
<\varepsilon/3$. Therefore, it follows that $\varepsilon=\|\varphi_1-\varphi_2\|_{T(O),j}=
\|\varphi_1-\psi+\psi-\varphi_2\|_{T(O),j} \leq \|\varphi_1-\psi\|_{T(O),j} +
\|\varphi_2-\psi\|_{T(O),j} < 2\varepsilon/3$, which is a contradiction. Hence,
${\mathfrak H}(T(O))$ is Hausdorff. The proof that $H(\oR^n;O)$ is a Hausdorff
locally convex space is immediate, by considering that the base for neighborhoods of
0 is generated by the open balls
\[
{\mathfrak B}_{i,n}(0)=\Bigl\{\varphi \in H(\oR^n;K_i) \mid
\|\varphi\|_{K_i,j}=\sup_{x \in \oR^n; \alpha \leq j}
\bigl\{e^{h_{K_i}(\xi)}|D^\alpha f(\xi)|\bigr\} < n^{-1}, n \in \oN\Bigr\}\,\,,
\]
and the proof is complete.
\end{proof}

\begin{theorem}
The spaces ${\mathfrak H}(T(O))$ and $H(\oR^n;O)$ are Fr\'echet spaces.
\label{Fspace}
\end{theorem}

\begin{proof}
That ${\mathfrak H}(T(O))$ is metrizable is clear from Theorem V.5 in~\cite{RS1},
if we endow the space ${\mathfrak H}(T(O))$ with the metric
$d(\varphi_1,\varphi_2)=\sum_{i=1}^\infty a_i \|\varphi_1-\varphi_2\|_{T(O),i}/
\bigl[1+\|\varphi_1-\varphi_2\|_{T(O),i}\bigr]$, such that $\sum_{i=1}^\infty a_i < \infty$.
Thus, it remains to show that ${\mathfrak H}(T(O))$ is complete. Let $\{\varphi_n\}$ be
a sequence of functions in ${\mathfrak H}(T(O))$. We shall take $\varphi_j \in \{\varphi_n\}$.
Given $\varepsilon > 0$, there exists $n_o$ such that for $p \geq n_o$ and $n \geq n_o$, we
have $d(\varphi_j,\varphi_n) < \varepsilon/2$ and $d(\varphi_j,\varphi_p) < \varepsilon/2$.
Then, it follows that $d(\varphi_p,\varphi_n) \leq d(\varphi_j,\varphi_p) +
d(\varphi_j,\varphi_n) < \varepsilon/2 + \varepsilon/2 = \varepsilon$. This proves that
$\{\varphi_n\}$ is Cauchy and hence ${\mathfrak H}(T(O))$ is complete. Thus
${\mathfrak H}(T(O))$ is Fr\'echet. For the proof that $H(\oR^n;O)$ is Fr\'echet
see~\cite{Mari1} (and in the case of $O=\oR^n$ see~\cite{Hasumi}). 
\end{proof}

It is an elementary fact that ${\mathfrak H}(T(O))$ and $H(\oR^n;O)$ are Banach spaces.

\begin{theorem}[Br\"uning-Nagamachi~\cite{BruNa1}, Proposition 2.6]
Let $O \subset \oR^n$ be a nonempty convex open subset. Then the spaces
${\mathfrak H}(T(O))$ and $H(\oR^n;O)$ are nuclear Fr\'echet spaces and,
in particular, reflexive.
\label{Nspace}
\end{theorem}

In light of the Theorems \ref{Fspace} and \ref{Nspace}, it follows that the spaces
${\mathfrak H}(T(O))$ and $H(\oR^n;O)$ are barreled~\cite[Corollary 1, p.347]{Treves} and
quasi-complete~\cite[p.354]{Treves}. According to Treves~\cite[Corollary 3, p.520]{Treves}
and Schaefer~\cite[exercise 19b, p.194]{Schaefer}, each quasi-complete barreled nuclear
space is a Montel space. Thus, one immediately arrives at

\begin{corollary}
The spaces ${\mathfrak H}(T(O))$ and $H(\oR^n;O)$ are Montel spaces.
\label{MSpace}
\end{corollary}

\begin{theorem}[\cite{Hasumi,Mari1,BruNa1}]
The space ${\mathscr D}({\oR^n})$ of all $C^\infty$-functions
on $\oR^n$ with compact support is dense in $H(\oR^n;K)$ and $H(\oR^n;O)$.
Moreover, the space $H(\oR^n;\oR^n)$ is dense in $H(\oR^n;O)$ and in
$H(\oR^n;K)$, and $H(\oR^m;\oR^m) \otimes H(\oR^n;\oR^n)$ is dense in
$H(\oR^{m+n};\oR^{m+n})$.
\label{theoINJ}
\end{theorem}

\begin{theorem}[Kernel theorem~\cite{BruNa1}]
Let $M$ be a separately continuous multilinear functional on $[{\mathfrak H}(T(\oR^4))]^n$.
Then there is a unique functional $F \in {\mathfrak H}^\prime(T(\oR^{4n}))$, for all
$f_i \in {\mathfrak H}(T(\oR^4))$, $i=1,\ldots,n$ such that
$M(f_1,\ldots,f_n)=F(f_1 \otimes \cdots \otimes f_n)$.
\label{KernelTheo} 
\end{theorem}

\begin{theorem}[\cite{Mari1,BruNa1}]
The space ${\mathfrak H}(T(\oR^n))$ is dense in ${\mathfrak H}(T(O))$ and
the space ${\mathfrak H}(T(\oR^{m+n}))$ is dense in ${\mathfrak H}(T(O))$.
\label{theoINJE}
\end{theorem}

From Theorem \ref{theoINJ} we have the following injections~\cite{Mari1}:
$H^\prime(\oR^n;K) \hookrightarrow H^\prime(\oR^n;\oR^n)
\hookrightarrow {\mathscr D}^\prime(\oR^n)$ and
$H^\prime(\oR^n;O) \hookrightarrow H^\prime(\oR^n;\oR^n)
\hookrightarrow {\mathscr D}^\prime(\oR^n)$.

\begin{definition}
The dual space $H^\prime(\oR^n;O)$ of $H(\oR^n;O)$ is the space of
distributions of exponential growth.
\end{definition}

A distribution $V \in H^\prime(\oR^n;O)$ may be expressed as a finite
order deri\-va\-ti\-ve of a continuous function of exponential growth
\[
V=D^\gamma_\xi[e^{h_K(\xi)}g(\xi)]\,\,,
\]
where $g(\xi)$ is a bounded continuous function. For $V \in
H^\prime(\oR^n;O)$ the follo\-wing result is known:

\begin{lemma}[\cite{Mari1}]
A distribution $V \in {\mathscr D}^\prime(\oR^n)$ belongs to $H^\prime(\oR^n;O)$
if and only if there exists a multi-index $\gamma$, a convex compact set $K \subset O$
and a bounded continuous function $g(\xi)$ such that
\[
V=D^\gamma_\xi[e^{h_K(\xi)}g(\xi)]\,\,.
\]
\label{lemmaMari}
\end{lemma}

For any element $U \in {\mathfrak H}^\prime$, its Fourier transform is
defined to be a distribution $V$ of exponential growth, such that the
Parseval-type relation $V(\varphi)=U(\psi)$, $\varphi \in H$,
$\psi={\mathscr F}[\varphi] \in {\mathfrak H}$, holds. In the same way,
the inverse Fourier transform of a distribution $V$ of exponential growth
is defined by the relation $U(\psi)=V(\varphi)$, $\psi \in {\mathfrak H}$,
$\varphi={\mathscr F}^{-1}[\psi] \in H$.

\begin{proposition}[\cite{Mari1}]
If $\varphi \in H(\oR^n;O)$, the Fourier transform of $\varphi$ belongs
to the space ${\mathfrak H}(T(O))$, for any open convex
nonempty set $O \subset \oR^n$. By the dual Fourier transform
$H^\prime(\oR^n;O)$ is topologically isomorphic with the space
${\mathfrak H}^\prime(T(-O))$.
\label{Propo1}
\end{proposition}

Let us now recall very briefly the basic definition of tempered ultrahyperfunctions.
These are defined as elements of a certain subspace of $Z^\prime$ of
ultradistributions of Gel'fand and Shilov which admit representations in terms of
analytic functions on the complement of some closed horizontal strip of the complex
space, and having polynomial growth on the complement of an open neighborhood of
that strip.

Let $\boldsymbol{{\mathscr H}_\omega}$ be the space of all
functions $f(z)$ such that ({\it i}) $f(z)$ is analytic for $\{z \in \oC^n \mid
|{\rm Im}\,z_1| > p, |{\rm Im}\,z_2| > p,\ldots,|{\rm Im}\,z_n| > p\}$,
({\it ii}) $f(z)/z^p$ is bounded continuous  in
$\{z\in \oC^n \mid |{\rm Im}\,z_1| \geqq p,|{\rm Im}\,z_2| \geqq p,
\ldots,|{\rm Im}\,z_n| \geqq p\}$, where $p=0,1,2,\ldots$ depends on $f(z)$
and ({\it iii}) $f(z)$ is bounded by a power of $z$, $|f(z)|\leq
{\boldsymbol{\sf C}}(1+|z|)^N$, where ${\boldsymbol{\sf C}}$ and $N$ depend on
$f(z)$. Define the {\em kernel} of the mapping $f:{\mathfrak H}(T(\oR^n))
\rightarrow \oC$ by $\boldsymbol{\Pi}$, as the set of all $z$-dependent
pseudo-polynomials, $z\in \oC^n$ (a pseudo-polynomial is a
function of $z$ of the form $\sum_s z_j^s G(z_1,...,z_{j-1},z_{j+1},...,z_n)$,
with $G(z_1,...,z_{j-1},z_{j+1},...,z_n) \in \boldsymbol{{\mathscr H}_\omega}$).
Then, $f(z) \in \boldsymbol{{\mathscr H}_\omega}$ belongs to the kernel
$\boldsymbol{\Pi}$ if and only if $f(\psi(x))=0$,
with $\psi(x) \in {\mathfrak H}(T(\oR^n))$ and $x={\rm Re}\,z$.
Consider the quotient space ${\mathscr U}=\boldsymbol{{\mathscr H}_\omega}
/\boldsymbol{\Pi}$. The set ${\mathscr U}$ is the space of tempered
ultrahyperfunctions. Thus, we have the

\begin{definition}
The space of tempered ultrahyperfunctions, denoted by ${\mathscr U}(\oR^n)$,
is the space of continuous linear functionals defined on ${\mathfrak H}(T(\oR^n))$. 
\label{UHF}
\end{definition}

In the following, we will put ${\mathfrak H}={\mathfrak H}(\oC^n)={\mathfrak H}(T(\oR^n))$
and the dual space of ${\mathfrak H}$ will be denoted by ${\mathfrak H}^\prime$.

\begin{theorem}[Hasumi~\cite{Hasumi}, Proposition 5] The space of
tempered ultrahyperfunctions ${\mathscr U}$ is algebraically isomorphic
to the space of generalized functions ${\mathfrak H}^\prime$.
\label{HasumiTheo}
\end{theorem}

\subsection{Tempered Ultrahyperfunctions Corresponding to a Proper
Convex Cone}
\label{Subsec}
Next, we consider tempered ultrahyperfunctions in a setting which includes
the results of~\cite{Tiao1,Hasumi,Mari1} as special cases, by considering
analytic functions in tubular radial domains~\cite{Carmi1,Carmi2,DanHenri,Daniel1}
and hence includes the important setting for quantum field theory of tube domains
over light cones. All the results below are taken from Refs.~\cite{DanHenri,Daniel1}
and hence the proofs will not be repeated. 

We start by introducing some terminology and simple facts concerning cones.
An open set $C \subset \oR^n$ is called a cone if $x \in C$ implies $\lambda x \in C$
for all $\lambda > 0$. Moreover, $C$ is an open connected cone if $C$ is a cone and
if $C$ is an open connected set. In the sequel, it will be sufficient to assume for
our purposes that the open connected cone $C$ in $\oR^n$ is an open convex cone
with vertex at the origin and {\it proper}, that is, it contains no any straight
line. A cone $C^\prime$ is called compact in $C$ --
we write $C^\prime \Subset C$ -- if the projection ${\sf pr}{\overline C^{\,\prime}}
\overset{\text{def}}{=}{\overline C^{\,\prime}} \cap S^{n-1} \subset
{\sf pr}C\overset{\text{def}}{=}C \cap S^{n-1}$, where $S^{n-1}$ is the unit
sphere in $\oR^n$. Being given a cone $C$ in $x$-space, we associate with $C$ a
closed convex cone $C^*$ in $\xi$-space which is the set $C^*=\bigl\{\xi \in \oR^n
\mid \langle \xi,x \rangle \geq 0, \forall x \in C \bigr\}$.
The cone $C^*$ is called the {\em dual cone} of $C$.
By $T(C)$ we will denote the set $\oR^n+iC \subset \oC^n$. If $C$ is
open and connected, $T(C)$ is called the tubular radial domain in $\oC^n$,
while if $C$ is only open $T(C)$ is referred to as a tubular cone. In the
former case we say that $f(z)$ has a boundary value $U=BV(f(z))$ in ${\mathfrak H}^\prime$
as $y \rightarrow 0$, $y \in C$ or $y \in C^\prime \Subset C$, respectively, if
for all $\psi \in {\mathfrak H}$ the limit
\[
\langle U, \psi \rangle=\lim_{{\substack{y \rightarrow 0 \\
y \in C~{\rm or}~C^\prime}}} \int_{\oR^n} f(x+iy)\psi(x) d^nx\,\,,
\]
exists. We will deal with tubes defined as the set of all points $z \in \oC^n$
such that
\[
T(C)=\Bigl\{x+iy \in \oC^n \mid
x \in \oR^n, y \in C, |y| < \delta \Bigr\}\,\,,
\]
where $\delta > 0$ is an arbitrary number.

An important example of tubular radial domain used in quantum field theory
is the tubular radial domain with the forward light-cone, $V_+$, as its basis
\[
V_+=\Bigl\{z \in \oC^n \mid {\rm Im}\,z_1 >
\Bigl(\sum_{i=2}^n {\rm Im}^2\,z_i \Bigr)^{\frac{1}{2}}, {\rm Im}\,z_1 > 0 \Bigr\}\,\,.  
\]

Let $C$ be an open convex cone, and let $C^\prime \Subset C$.
Let $B[0;r]$ denote a {\bf closed} ball of the
origin in $\oR^n$ of radius $r$, where $r$ is an arbitrary positive
real number. Denote $T(C^\prime;r)=\oR^n+i\bigl(C^\prime \setminus
\bigl(C^\prime \cap B[0;r]\bigr)\bigr)$.
We are going to introduce a space of holomorphic functions
which satisfy certain estimate according to Carmichael~\cite{Carmi1}. We want to
consider the space consisting of holomorphic functions $f(z)$ such that
\begin{equation}
\bigl|f(z)\bigr|\leq {\boldsymbol{\sf C}}(C^\prime)(1+|z|)^N e^{h_{C^*}(y)}\,\,,\quad
z \in T(C^\prime;r)\,\,,
\label{eq31} 
\end{equation}
where $h_{C^*}(y)=\sup_{\xi \in C^*}|\langle \xi,y \rangle|$
is the indicator of $C^*$, ${\boldsymbol{\sf C}}(C^\prime)$ is a constant that depends
on an arbitrary compact cone $C^\prime$ and $N$ is a non-negative real number.
The set of all functions $f(z)$ which are holomorphic in $T(C^\prime;r)$ and
satisfy the estimate (\ref{eq31}) will be denoted by $\boldsymbol{{\mathscr H}^o_c}$.
Throughout the remainder of this paper $T(C^\prime;r)$ will
denote the set $\oR^n+i\bigl(C^\prime \setminus \bigl(C^\prime \cap B[0;r]\bigr)\bigr)$.

\begin{remark}
The space of functions $\boldsymbol{{\mathscr H}^o_c}$ constitutes a generalization
of the space ${\mathfrak A}_{_\omega}^i$ of Sebasti\~ao e Silva~\cite{Tiao1} and the
space $\mona_{_\omega}$ of Hasumi~\cite{Hasumi} to arbitrary tubular radial domains
in $\oC^n$.
\end{remark}

\begin{lemma}[\cite{Carmi2,DanHenri}]
Let $C$ be an open convex cone, and let $C^\prime \Subset C$.
Let $h(\xi)=e^{k|\xi|}g(\xi)$, $\xi \in \oR^n$, be
a function with support in $C^*$, where $g(\xi)$ is a bounded continuous
function on $\oR^n$. Let $y$ be an arbitrary but fixed point of
$C^\prime \setminus \bigl(C^\prime \cap B[0;r]\bigr)$. Then
$e^{-\langle \xi,y \rangle}h(\xi) \in L^2$, as a function of $\xi \in \oR^n$.
\label{lemma0}
\end{lemma}

\begin{definition}
We denote by $H^\prime_{C^*}(\oR^n;O)$ the subspace of $H^\prime(\oR^n;O)$
of distributions of exponential growth with support in the cone $C^*$:
\begin{equation}
H^\prime_{C^*}(\oR^n;O)=\Bigl\{V \in H^\prime(\oR^n;O) \mid
\supp(V) \subseteq C^* \Bigr\}\,\,. 
\label{eq31'} 
\end{equation}
\end{definition}

\begin{lemma}[\cite{Carmi2,DanHenri}]
Let $C$ be an open convex cone, and let $C^\prime \Subset C$.
Let $V=D^\gamma_\xi[e^{h_K(\xi)}g(\xi)]$, where
$g(\xi)$ is a bounded continuous function on $\oR^n$ and $h_K(\xi)=k|\xi|$
for a convex compact set $K=\bigl[-k,k\bigr]^n$. Let
$V \in H^\prime_{C^*}(\oR^n;O)$. Then $f(z)=(2\pi)^{-n}
(V,e^{-i\langle \xi,z \rangle})$ is an element of
$\boldsymbol{{\mathscr H}^o_c}$.
\label{lemma1}
\end{lemma}

We now shall define the main space of holomorphic functions with which this paper
is concerned. Let $C$ be a proper open convex cone, and let $C^\prime \Subset C$.
Let $B(0;r)$ denote an {\bf open} ball of the origin in $\oR^n$ of radius
$r$, where $r$ is an arbitrary positive real number. Denote $T(C^\prime;r)=
\oR^n+i\bigl(C^\prime \setminus \bigl(C^\prime \cap B(0;r)\bigr)\bigr)$. Throughout
this section, we consider functions $f(z)$ which are holomorphic in
$T(C^\prime)=\oR^n+iC^\prime$ and which satisfy the estimate (\ref{eq31}),
with $B[0;r]$ replaced by $B(0;r)$. We denote this space by
$\boldsymbol{{\mathscr H}^{*\,o}_c}$. We note that $\boldsymbol{{\mathscr H}^{*\,o}_c}
\subset \boldsymbol{{\mathscr H}^{o}_c}$ for any open convex cone $C$. Put
${\mathscr U}_c=\boldsymbol{{\mathscr H}^{*\,o}_c}/\boldsymbol{\Pi}$, that is,
${\mathscr U}_c$ is the quotient space of $\boldsymbol{{\mathscr H}^{*\,o}_c}$
by set of pseudo-polynomials $\boldsymbol{\Pi}$.

\begin{definition}
The set ${\mathscr U}_c$ is the space of tempered ultrahyperfunctions corresponding
to a proper open convex cone $C \subset \oR^n$.
\end{definition}

The following theorem shows that functions in $\boldsymbol{{\mathscr H}^{*\,o}_c}$
have distributional boundary values in ${\mathfrak H}^\prime$. Further, it shows
that functions in $\boldsymbol{{\mathscr H}^{*\,o}_c}$ satisfy a strong boundedness
property in ${\mathfrak H}^\prime$.

\begin{theorem}[\cite{Daniel1}]
Let $C$ be an open convex cone, and let $C^\prime \Subset C$.
Let $V=D^\gamma_\xi[e^{h_K(\xi)}g(\xi)]$, where $g(\xi)$ is a bounded continuous
function on $\oR^n$ and $h_K(\xi)=k|\xi|$ for a convex compact set
$K=\bigl[-k,k\bigr]^n$. Let $V \in H^\prime_{C^*}(\oR^n;O)$. Then

\,\,\,$(i)\quad f(z)=(2\pi)^{-n}(V,e^{-i\langle \xi,z \rangle})$
is an element of $\boldsymbol{{\mathscr H}^{*\,o}_c}$,

\,\,\,$(ii)\quad \bigl\{f(z) \mid y={\rm Im}\,z \in C^\prime \Subset C, |y| \leq Q\bigr\}$
is a strongly bounded set in ${\mathfrak H}^\prime$, where $Q$ is an arbitrarily but fixed
positive real number,

\,\,\,$(iii)\quad f(z) \rightarrow {\mathscr F}^{-1}[V] \in {\mathfrak H}^\prime$ in the
strong (and weak) topology of ${\mathfrak H}^\prime$ as $y={\rm Im}\,z \rightarrow 0$,
$y \in C^\prime \Subset C$.
\label{theorem1}
\end{theorem}

The functions $f(z) \in \boldsymbol{{\mathscr H}^{*\,o}_c}$ can be recovered as
the (inverse) Fourier-Laplace transform of the constructed distribution $V \in
H^\prime_{C^*}(\oR^n;O)$. This result is a generalization of the Paley-Wiener-Schwartz
theorem for the setting of tempered ultrahyperfunctions.

\begin{theorem}[Paley-Wiener-Schwartz-type Theorem~\cite{Daniel1}]
Let $f(z) \in \boldsymbol{{\mathscr H}^{*\,o}_c}$, where $C$ is an open convex cone.
Then the distribution $V \in H^\prime_{C^*}(\oR^n;O)$ has a uniquely
determined inverse Fourier-Laplace transform $f(z)=(2\pi)^{-n}(V,e^{-i\langle \xi,z \rangle})$
which is holomorphic in $T(C^\prime)$ and satisfies the estimate (\ref{eq31}), with $B[0;r]$
replaced by $B(0;r)$.
\label{PWSTheo} 
\end{theorem}

The following corollary is immediate from Theorem \ref{PWSTheo}.

\begin{corollary}[\cite{BruNa1}]
Let $C^*$ be a closed convex cone and $K$ a convex compact set in
$\oR^n$. Define an indicator function $h_{K,C^*}(y)$, $y \in \oR^n$,
and an open convex cone $C_K$ such that
$h_{K,C^*}(y)=\sup_{\xi \in C^*}\bigl|h_K(\xi)-\langle \xi,y \rangle\bigr|$
and $C_K=\bigl\{y \in \oR^n \mid h_{K,C^*}(y)\!<\!\infty\bigr\}$.
Then the distribution $V \in H^\prime_{C^*}(\oR^n;O)$ has a uniquely
determined inverse Fourier-Laplace transform $f(z)=(2\pi)^{-n}
(V,e^{-i\langle \xi,z \rangle})$ which is holomorphic in the tube
$T(C^\prime_K)=\oR^n+iC^\prime_K$, and satisfies the following estimate,
for a suitable $K \subset \oR^n$,
\begin{equation}
\bigl|f(z)\bigr|\leq {\boldsymbol{\sf C}}(C^\prime)(1+|z|)^N e^{h_{K,C^*}(y)}\,\,,
\quad z \in T(C^\prime_K;r)=\oR^n+i\bigl(C^\prime_K \setminus \bigl(C^\prime_K
\cap B(0;r)\bigr)\bigr)
\label{eq37} 
\end{equation}
where $C^\prime_K \Subset C_K$.
\label{PWSCorol}
\end{corollary}

The same proof as in Carmichael~\cite[Theorem 1, equation (4)]{Carmi2}
combined with the proofs of Theorems \ref{theorem1} and \ref{PWSTheo} shows that
the following theorem is true.

\begin{theorem}
Let $C$ be an open convex cone, and let $C^\prime \Subset C$.
Let $f(z) \in \boldsymbol{{\mathscr H}^{*\,o}_c}$. Then there exists
a unique element $V \in H^\prime_{C^*}(\oR^n;O)$ such that
\begin{equation}
{f(z)}={\mathscr F}^{-1}\bigl[e^{-\langle \xi,y \rangle}
V \bigr]\,\,,\quad z \in T(C^\prime;r)=\oR^n+i\bigl(C^\prime \setminus
\bigl(C^\prime \cap B(0;r)\bigr)\bigr)\,\,,
\label{equa12}
\end{equation}
where (\ref{equa12}) holds as an equality in ${\mathfrak H}^\prime(T(O))$.  
\label{theol2}
\end{theorem}

\begin{remark}
It is important to remark that in Theorems \ref{theorem1} and \ref{PWSTheo}
we are considering the inverse Fourier-Laplace transform $f(z)=(2\pi)^{-n}
\bigl\langle V,e^{-i\langle \xi,z \rangle} \bigr\rangle$, in opposition to the
Fourier-Laplace transform used in the proof of Theorem 1 of Ref.~\cite{Carmi2}.
In this case the proof of Theorem \ref{theol2} is achieved if we consider
$\xi$ as belon\-ging to the open half-space $\bigl\{\xi \in C^* \mid \langle \xi,y
\rangle < 0\bigr\}$, for $y \in C^\prime \setminus \bigl(C^\prime \cap B(0;r)\bigr)$,
since by hypothesis $f(z) \in \boldsymbol{{\mathscr H}^{*\,o}_c}$. Then,
from~\cite[Lemma 2, p.223]{Vlad} there is $\delta(C^\prime)$ such that for
$y \in C^\prime \setminus \bigl(C^\prime \cap B(0;r)\bigr)$ implies $\langle
\xi,y \rangle \leq -\delta(C^\prime) |\xi||y|$. This justifies the negative sign
in (\ref{equa12}).
\label{remark2}
\end{remark}

In this point, we note the following fact important. Let ${\mathfrak H}_C^\prime(T(O))$
denote the subset of ${\mathfrak H}^\prime(T(O))$ defined by ${\mathfrak H}_C^\prime(T(O))
=\bigl\{U \in {\mathfrak H}^\prime(T(O)) \mid U={\mathscr F}[V], V \in
H^\prime_{C^*}(\oR^n;O)\bigr\}$. Then, by exactly the same arguments explained
in~\cite[p.114]{Carmi3}, we have the following corollary of Theorems \ref{theorem1},
\ref{PWSTheo} and \ref{theol2}.

\begin{corollary}
Let $C$ be an open convex cone. Then $\boldsymbol{{\mathscr H}^{*\,o}_c}$ is
algebraically isomorphic to both $H^\prime_{C^*}(\oR^n;O)$ and ${\mathfrak H}_C^\prime(T(O))$. 
\end{corollary}

We finish this section with two results proved in Ref.~\cite{Daniel1}, which
will be used in the applications of Section \ref{SecTheorems}.

\begin{theorem}[Ultrahyperfunctional version of edge of the wedge
theorem]
Let $C$ be an open cone of the form $C=C_1 \cup C_2$, where each $C_j$, $j=1,2$,
is a proper open convex cone. Denote by $\boldsymbol{ch}(C)$ the convex hull of the
cone $C$. Assume that the distributional boundary values of two holomorphic functions
$f_j(z) \in \boldsymbol{{\mathscr H}^{*\,o}_{c_j}}$ $(j=1,2)$ agree, that
is, $U=BV(f_1(z))=BV(f_2(z))$, where $U \in {\mathfrak H}^\prime$ in accordance with
the Theorem \ref{theorem1}. Then there exists $F(z) \in
\boldsymbol{{\mathscr H}^{o}_{{\boldsymbol{ch}(C)}}}$
such that $F(z)=f_j(z)$ on the domain of definition of each $f_j(z)$, $j=1,2$.
\label{EWTheo}
\end{theorem}

\begin{theorem}
Let $C$ be some open convex cone. Let $f(z) \in \boldsymbol{{\mathscr H}^{*\,o}_{c}}$.
If the boundary value $BV(f(z))$ of $f(z)$ in the sense of tempered ultrahyperfunctions
vanishes, then the function $f(z)$ itself vanishes.
\label{UniTheo}
\end{theorem}

\section{Wightman Functionals for UHFNCQFT and Their Properties}
\label{SecAxioms}
According to Wightman, the conventional postulates of QFT can be fully reexpressed
in terms of an equivalent set of properties of the vacuum expectation values
of their ordinary field products, called Wightman distributions
\begin{eqnarray}
{\mathfrak W}_m(f_{1} \otimes \cdots \otimes f_{m})\overset{\text{\rm def}}
{=}\langle\Omega_o \mid \Phi(f_1) \cdots \Phi(f_m) \mid \Omega_o\rangle\,\,,
\label{vev}
\end{eqnarray}
where $(f_{1} \otimes \cdots \otimes f_{m})= f_1(x_1) \cdots f_m(x_m)$
is considered as an element of ${\mathscr S}(\oR^{4m})$, and
$\mid \Omega_o\rangle$ is the vacuum vector, unique vector time-translation
invariant of the Hilbert space of states.

\begin{remark}
To keep things as simple as possible, we will assume that the Wightman
distributions are ``functions'' ${\mathfrak W}_m(x_{1},\ldots,x_{m})$.
The reader can easily supply the necessary test functions.
\end{remark}

As a general rule, the continuous
linear functionals ${\mathfrak W_m}(x_{1},\ldots,x_{m})$ are assumed to satisfy
the following properties:

\newcounter{numero}
\setcounter{numero}{0}
\def\Prop{\addtocounter{numero}{1}\item[{$\boldsymbol{\sf P_{\thenumero}}$}]}
\begin{enumerate}

\Prop (Temperedness). The sequence of Wightman functions
${\mathfrak W_m}(x_{1},\ldots,x_{m})$ are tempered distributions in
${\mathscr S}^\prime(\oR^{4m})$, for all $m \geq 1$. This property is included
in the list of properties for a QFT for technical reasons.

\bigskip

\Prop (Poincar\'e Invariance). Wightman functions are invariant under the
Poincar\'e group
\[
{\mathfrak W_m}(\Lambda x_{1}+a,\ldots,\Lambda x_{m}+a)=
{\mathfrak W_m}(x_{1},\ldots,x_{m})\,\,.
\]

\bigskip

\Prop (Spectral Condition). The Fourier transforms of the Wightman functions
have support in the region
\begin{gather*}
\Bigm\{(p_1,\ldots,p_m)\in {\Bbb R}^{4m}\,\,\bigm|\,\,\sum_{j=1}^{m}p_j=0,\,\,
\sum_{j=1}^{k}p_j \in {\overline V}_+,\,\,k=1,\dots,m-1 \Bigm\}\,\,,
\tag{SC}
\end{gather*}
where ${\overline V}_+=\{(p^0,{\boldsymbol p}) \in \oR^4 \mid p^2 \geq 0,
p^0 \geq 0\}$ is the closed forward light cone.

\bigskip

\Prop (Local commutativity). This property has origin in the quantum
principle that operator observables $\Phi(x)$ corresponding to independent
measurements must comute. 
\[
{\mathfrak W}_m(x_{1},\ldots,x_j,x_{j+1},\ldots,x_{m})=
{\mathfrak W}_m(x_{1},\ldots,x_{j+1},x_j,\ldots,x_{m})\,\,,
\]
if $(x_j-x_{j+1})^2<0$.

\bigskip
 
\Prop For any finite set $f_o,f_1,\ldots,f_N$ of test functions such that
$f_o \in \oC$, $f_j \in {\mathscr S}(\oR^{4j})$ for $1 \leq j \leq N$,
one has
\begin{align*}
\sum_{k,\ell=0}^N {\mathfrak W}_{k+\ell}(f_k^* \otimes f_\ell)
\geq 0\,\,.
\end{align*}

\bigskip

\Prop (Hermiticity). A neutral scalar field must be real valued. This implies that
\[
{\mathfrak W}_m(x_{1},x_2,\ldots,x_{m-1},x_{m})=
\overline{{\mathfrak W}_m(x_{m},x_{m-1},\ldots,x_{1},x_2})\,\,.
\]

\end{enumerate} 

Generalizing these properties to NCQFT is not as simple, especially the
Lorentz symmetry. For example, as already mentioned in the
Introduction, the Lorentz symmetry is not preserved in NCQFT. Furthemore,
the existence of hard infrared singularities in the non-planar sector of
the theory can destroy the {\em tempered} nature of the Wightman functions.
And more, how can the Property $\boldsymbol{\sf P_4}$ be described in
field theory with a fundamental length? In order to answer these questions,
we shall assume a NCFT where the Wightman functionals fulfil a set of properties
which actually will characterize a UHFNCQFT.

\subsection{Twisted Poincar\'e Symmetry}
\label{Subsec1}
In this paper, we will assume that our fields are transforming according
to representations of the {\em twisted} Poincar\'e group~\cite{Chai,ChaTu}.
This formalism has the advantage of retaining the Wigner's notion of
elementary particles.\footnote{Another approach where the full Poincar\'e
group is preserved was proposed by Doplicher-Fredenhagen-Roberts~\cite{DFR}.}  

When referring to NCQFT one should have in mind the deformation of the
ordinary product of fields. This deformation is performed in terms of the
star product extended for noncoinciding points via the functorial
relation~\cite{Szabo}
\begin{align}
\varphi(x_1)\star &\cdots \star \varphi(x_n)=\prod\limits_{i<j}
\exp \left({\frac{i}{2}\theta^{\mu\nu}\frac{\partial}{\partial x_i^\mu}
\otimes \frac{\partial}{\partial x_j^\nu}} \right)\varphi(x_1)\cdots \varphi(x_n)\,\,.
\label{Moyalprod}
\end{align}
For coinciding points $x_1=x_2=\cdots=x_n$ the product (\ref{Moyalprod}) becomes identical
to the multiple Moyal $\star$-product. We shall consider NCQFT in the sense of a field theory
on a non-commutative spacetime encoded by a Moyal product on the test function algebra.

\begin{definition}[Vacuum Expectation Values of Fields~\cite{Chai1}]
In a UHFNCQFT the Wightman functionals in ${\mathscr U}_c(\oR^{4m})$,
i.e., the $m$-points vacuum expectation values of fields operators
are defined by
\begin{eqnarray}
{\mathfrak W}_m^{\star}(z_{1},\ldots,z_{m})\overset{\text{\rm def}}
{=}\langle\Omega_o \mid \Phi(z_1)\star \cdots \star \Phi(z_m) \mid \Omega_o\rangle\,\,.
\label{NWF}
\end{eqnarray}
\end{definition}

\begin{remark}
The tempered ultrahyperfunctions ${\mathfrak W}^\star_m \in {\mathscr U}_c(\oR^{4m})$
will be called {\em non-commutative} Wightman functions.
\end{remark}

\begin{remark}
In~\cite{Vernov} the Wightman functions were written as follows:
\[
{\mathfrak W}_m^{\tilde\star}(z_{1},\ldots,z_{m})\overset{\text{\rm def}}
{=}\langle\Omega_o \mid \Phi(z_1)\tilde\star \cdots \tilde\star\, \Phi(z_m)
\mid \Omega_o\rangle\,\,,
\]
where the meaning of $\tilde\star$ depends on the considered case. In particular,
if $\tilde\star = 1$, we obtain the standard form ${\mathfrak W}_m(z_{1},\ldots,z_{m})=
\langle\Omega_o \mid \Phi(z_1) \cdots \Phi(z_m) \mid \Omega_o\rangle$ adopted in~\cite{AGVM},
which corresponds to the commutative theory with the $SO(1,1) \times SO(2)$ invariance.
On the other hand, if $\tilde\star = \star$, this choice corresponds to the Wightman functions
introduced in~\cite{Chai1}. In this case, the non-commutativity is manifested not only at
coincident points but also in their neighborhood.
\end{remark}

As a consequence of the twisted Poincar\'e covariance condition of the $\star$-product of
fields~\cite{ChaTu}, the non-commutative Wightman functions ${\mathfrak W}^\star_m(z_1,\ldots,z_m)
\in {\mathscr U}_c(\oR^{4m})$ sa\-tis\-fy the twisted Poincar\'e transformations (besides of the
symmetry $SO(1,1) \times SO(2)$). Thus, we have the

\begin{theorem} ${\mathfrak W}^\star_m(z_1,\ldots,z_m)=
{\mathfrak W}^\star_m(\Lambda z_1+a,\ldots,\Lambda z_m+a)$, in the usual distributional
sense.
\end{theorem}

\subsection{Domain of Analyticity of Non-Commutative Wightman Functions}
\label{Subsec3}
Since for non-commutative theories the group of translations is intact, the
Wightman functions only depends on the $(m-1)$ coordinate differences as in the
commutative case. Then, passing to the difference variables $\zeta_i$,
we obtain, symbolically, that
\[
{\mathfrak W}^\star_m(z_1,\ldots,z_m)=W^\star_m(\zeta_1,\ldots,\zeta_{m-1})\,\,,\quad
\zeta_j=z_j-z_{j+1}\,\,,\quad j=1,\ldots,m-1\,\,.
\]

Applying Corollary \ref{PWSCorol} to the ordinary Wightman functions
$W_m(\zeta_1,\ldots,\zeta_{m-1})$, we obtain the following important result:

\begin{theorem}
The functions $W_{m-1}(\zeta_1,\ldots,\zeta_{m-1})$ are holomorphic functions
of $4(m-1)$ complex variables in a set which contains $\oR^{4(m-1)}+
V_+(\ell_{\theta_1},\ldots,\ell_{\theta_{m-1}})$, where
\[
V_+(\ell_{\theta_1},\ldots,\ell_{\theta_{m-1}})=
\Bigl\{(\eta_1,\ldots,\eta_{m-1}) \in \oR^{4(m-1)} \mid \eta_j=y_j +
(\ell_{\theta_j}, {\boldsymbol 0}) \in V_+
+ (\ell_{\theta_j}, {\boldsymbol 0}) \Bigr\}\,,
\]
and satisfy the estimate
\begin{equation}
\bigl|W_{m-1}(\zeta_1,\ldots,\zeta_{m-1})\bigr|\leq {\boldsymbol{\sf C}}(V^\prime)
\prod_{j=1}^{m-1}(1+|\zeta_j|)^N
\exp\Bigl({h_{K,\overline{V}_+^{m-1}}(y_j)}\Bigr)\,\,.
\label{eq37*} 
\end{equation}
\end{theorem}

\begin{proof}
The first part of theorem follows immediately from Remark 2.18 in~\cite{BruNa1}.
Thus we need only show that $W_{m-1}(\zeta_1,\ldots,\zeta_{m-1})$ satisfies the estimate
(\ref{eq37*}). But, this can be proved by using the Theorem \ref{PWSTheo}
in order to show that the function $W_{m-1}(\zeta_1,\ldots,\zeta_{j-1},\zeta^\prime,
\zeta_{j+1},\ldots,\zeta_{m-1})$ is a holomorphic function of $\zeta^\prime$ alone, with the
complex variables $\zeta_1,\ldots,\zeta_{j-1},\zeta_{j+1},\ldots,\zeta_{m-1}$ being kept fixed.
Then, we apply this argument, in turn, to each variable $\zeta_j$ separately.
\end{proof}

\begin{proposition}
In a UHFNCQFT the Wightman functionals in ${\mathscr U}_c(\oR^{4(m-1)})$,
i.e., the non-commutative Wightman functions involving the $\star$-product,
$W^\star_{m-1}$, coincide with the standard Wightman functions $W_{m-1}$.
\label{PropEquiv}
\end{proposition}

\begin{proof}
By considering that in terms of complex variables
\[
\prod\limits_{i<j}
\exp \left({\frac{i}{2}\theta^{\mu\nu}\frac{\partial}{\partial x_i^\mu}
\otimes \frac{\partial}{\partial x_j^\nu}} \right)=
\prod\limits_{i<j}
\exp \left({\frac{1}{2}\theta^{\mu\nu}\frac{\partial}{\partial \zeta_i^\mu}
\wedge \frac{\partial}{\partial \bar{\zeta}_j^\nu}} \right)\,\,,
\]
and since the functions $W_m(\zeta_1,\ldots,\zeta_{m-1})$ are holomorphic,
then it follows that
\[
W^\star_{m-1}(\zeta_1,\ldots,\zeta_{m-1})=W_{m-1}(\zeta_1,\ldots,\zeta_{m-1})\,\,,
\]
and the proof is complete.
\end{proof}

\begin{corollary}
The non-commutative Wightman functions $W^\star_{m-1}(\zeta_1,\ldots,\zeta_{m-1})$ are
holomorphic functions of $4(m-1)$ complex variables in a set which contains $\oR^{4(m-1)}+
V_+(\ell_{\theta_1},\ldots,\ell_{\theta_{m-1}})$, and satisfy the estimate
\begin{equation*}
\bigl|W^\star_{m-1}(\zeta_1,\ldots,\zeta_{m-1})\bigr|\leq {\boldsymbol{\sf C}}(V^\prime)
\prod_{j=1}^{m-1}(1+|\zeta_j|)^N
\exp\Bigl({h_{K,\overline{V}_+^{m-1}}(y_j)}\Bigr)\,\,. 
\end{equation*}
\label{Corol45}
\end{corollary}

It is suggestive to see that $W^\star_{m-1}$ has the same form as the standard form
$W_{m-1}$ in a UHFNCQFT. In light of Proposition \ref{PropEquiv}, where we have as result
that $\tilde\star=\star=1$, we conjecture that the possibility of extending the axiomatic
approach to the NCQFT in terms of tempered ultrahyperfunctions is {\it independent} of the
concrete type of the $\tilde\star$-product (similar conclusion was obtained in~\cite{Vernov}).
In order to support this conjecture, in Section \ref{SecTheorems}, we derive for the UHFNCQFT
the validity of some important theorems. These include the existence of CPT symmetry and the
connection between Spin and Statistics for UHFNCQFT, in the case of space-space non-commutativity.
In what follows, we shall always refer to the functions $W^\star_{m-1}$ in order to include
non-commutativity effects not only into the vacuum state, as it happens with the functions $W_{m-1}$. 

\subsection{Extended Local Commutativity Condition}
\label{subsec2}
The existence of a minimum length related to the scale of nonlocality
$\ell_\theta$~\cite{Szabo} renders impossible the preservation of the canonical
commutation rules since those rules make sense only in the distance regions greater
than $\ell_\theta$. Thus, in order to remedy this difficulty the local commutativity
will be replaced by a distinguished localization property in the sense of
Br\"uning-Nagamachi~\cite{BruNa1}, called {\em extended local commutativity}. This
property is defined as a continuity condition of the expectation values of the field
commutators in a topology associated to a $\ell_\theta$-neighborhood of the light cone.

Let $|x|_1$ be the norm
\[
|x|_1=|x_0|+|{\boldsymbol x}|\,\,,\quad
|{\boldsymbol x}|=\sqrt{\sum_{i=1}^3\,\,(x_j)^2}\,\,,
\]
for $x=(x_0,{\boldsymbol x}) \in \oR^4$. Denote
\[
L^{\ell}=\Bigl\{(x_1,x_2) \in \oR^8 \mid |x_1-x_2|_1 < \ell_\theta \Bigr\}\,\,.
\]
Define the open set $V_+$ of all strictly time-like points in $\oR^4$ by
\[
V_+=\Bigl\{x \in \oR^4 \mid (x_0)^2 - {\boldsymbol x}^2 > 0 \Bigr\}\,\,.
\]
In order to prepare for the definition of the extended local commutativity,
we shall consider functionals which are carried by sets
close to $\oR^4$ but not contained in $\oR^4$. Denote by $V^{\ell_\theta}$
the complex $\ell_\theta$-neighborhood of $V_+$
\[
V^{\ell_\theta}=\Bigl\{z \in \oC^4 \mid
\exists\,\,x \in V_+, |{\rm Re}\,z - x| +
|{\rm Im}\,z|_1 < \ell_\theta \Bigr\}\,\,.
\]
Consider the set of all pairs of
points in $\oC^4$ whose difference belongs to the $\ell_\theta$-neighborhood, 
\[
M^{\ell_\theta}=\Bigl\{(z_1,z_2) \in \oC^8 \mid
z_1-z_2 \in V^{\ell_\theta} \Bigr\}\,\,,
\]
and introduce the space ${\mathfrak H}(M^{\ell_\theta})$ consisting of all
holomorphic functions on $M^{\ell_\theta}$. Then, according to
Br\"uning-Nagamachi~\cite{BruNa1}, we formulate the
axiom of extended local commutativity condition as follows.

\begin{definition}[extended local commutativity condition]
Let $f,g$ be two test functions in ${\mathfrak H}(T(\oR^4))$, then
the fields $\Phi(f)$ and $\Phi(g)$ are said to commute for any
relative spatial separation $\ell^\prime > \ell_\theta$ of their arguments,
if the functional
\begin{align}
{\boldsymbol{\sf F}}&=\bigl\langle\Theta \mid \bigl[\varphi(f),\varphi(g)\bigr]_\star
\mid \Psi\bigr\rangle \nonumber\\[3mm]
&=\bigl\langle\Theta \mid \bigl(\varphi(f) \star \varphi(g)-
\varphi(g) \star \varphi(f)\bigr)\mid \Psi\bigr\rangle\,\,,
\label{AofAC}
\end{align}
is carried by the set $M^{\ell^\prime}=\bigl\{\bigr(z_1,z_2)
\in \oC^{8} \mid z_1-z_2 \in V^{\ell^\prime}\}$, for any vectors
$\Theta,\Psi\in D_0$, i.e., if the functional ${\boldsymbol{\sf F}}$ can be extended
to a continuous linear functional on ${\mathfrak H}(M^{\ell^\prime})$.
\label{ELC}
\end{definition}

The Definition \ref{ELC} can be understood saying that two operators $\Phi(f)$
and $\Phi(g)$, at two distinct points of the non-commutative spacetime,
can not be distinguished if the relative spatial distance between their arguments
is less than $\ell_\theta$. In other words, in NCQFT the quantum fluctuations of the
spacetime {\em operationally} prevent the exact localization of the events inside of
the minimum area $\ell_\theta^2$. This area is interpreted
as the minimum region which {\em observables} can be probed~\cite{Calmet1}.

Moreover, it follows from the extended local commutativity condition and from
the Propositions 4.3 and 4.4 in~\cite{BruNa1} that the functional
${\boldsymbol{\sf F}} \in {\mathscr U}_c(\oR^{4m})$ defined by
\begin{align*}
{\boldsymbol{\sf F}}={\mathfrak W}^\star_{m}(z_1,\ldots,z_j,z_{j+1},\ldots,z_m)
-{\mathfrak W}^\star_{m}(z_1,\ldots,z_{j+1},z_j,\ldots,z_m)\,\,,
\end{align*}
for any $\ell^\prime > \ell_\theta$, $m \geq 2$ and $j \in \{1,\ldots,m-1\}$, can
be extended to a continuous linear functional on ${\mathfrak H}(M_j^{\ell^\prime})$,
with $M_j^{\ell^\prime}=\Bigl\{\bigr(z_1,\ldots,z_m) \in \oC^{4m} \mid z_j-z_{j+1}
\in V^{\ell^\prime}\Bigr\}$.

\subsection{Properties of Non-Commutative Wightman Functions}
\label{Subsec5}
The analysis of the preceding results has shown that the sequence of vacuum
expectation values of a NCQFT in terms of tempered ultrahyperfunctions satisfies
a number of specific properties. We summarize these below:

\setcounter{numero}{0}
\def\Prop{\addtocounter{numero}{1}\item[{$\boldsymbol{\sf P^\prime_{\thenumero}}$}]}
\begin{enumerate}

\Prop ${\mathfrak W^\star_0}=1$, ${\mathfrak W^\star_m} \in {\mathscr U}_c(\oR^{4m})$
for $n \geq 1$, and ${\mathfrak W^\star_m}(f^*)=\overline{{\mathfrak W^\star_m}(f)}$,
for all $f \in {\mathfrak H}(T(\oR^{4m}))$, where $f^*(z_1,\ldots,z_m)=
\overline{f(\bar{z}_1,\ldots,\bar{z}_m)}$.

\bigskip

\Prop The Wightman functionals ${\mathfrak W^\star_m}$ are invariant under
the {\em twisted} Poincar\'e group

\bigskip

\Prop Spectral condition. Since the Fourier transformation of tempered
ultrahyperfunctions are distributions, the spectral condition is not
so much different from that of Schwartz distributions. Thus, for every
$m \in \oN$, there is $\widehat{{\mathfrak W}}^\star_{m} \in
H_{V^*}^\prime(\oR^{4m},\oR^{4m})$~\cite{BruNa1}, where
\begin{equation}
H^\prime_{V^*}(\oR^{4m},\oR^{4m})=\Bigl\{V \in H^\prime(\oR^{4m},\oR^{4m}) \mid
\supp\,(\widehat{{\mathfrak W}}^\star_{m}) \subset V^* \Bigr\}\,\,, 
\label{EQ31'} 
\end{equation}
with $V^*$ being the properly convex cone (SC) defined in $\boldsymbol{\sf P_3}$.

\bigskip

\Prop Extended local commutativity condition.

\bigskip
 
\Prop For any finite set $f_o,f_1,\ldots,f_N$ of test functions such that
$f_o \in \oC$, $f_j \in {\mathfrak H}(T(\oR^{4j}))$ for $1 \leq j \leq N$,
one has
\begin{align*}
\sum_{k,\ell=0}^N {\mathfrak W}^\star_{k+\ell}(f_k^* \otimes f_\ell)
\geq 0\,\,.
\end{align*}

\end{enumerate} 

\section{CPT, Spin-Statistics and All That in UHFNCQFT}
\label{SecTheorems}
In the preceding sections, we have defined what is meant by NCQFT in terms
of tempered ultrahyperfunctions and assembled some tools to aid in the
analysis of its structure. In this section, these are used to establish
some important theorems as the celebrated CPT and spin-statistics
theorems. The proof of these results as given in the
literature~\cite{SW}-\cite{Haag} usually seem to rely on the local
character of the distributions in an essential way. In the approach which we follow the
apparent source of difficulties in proving these results is the fact that for
functionals belonging to the space of tempered ultrahyperfunctions the standard notion
of the localization principle breaks down.

Let $\Phi$ be a Hermitian scalar field. For this field, it is well-known that in terms
of the Wightman functions, a necessary and sufficient condition for the existence of CPT
theorem is given by:
\begin{equation}
{\mathfrak W}_m(x_1,\ldots,x_m)={\mathfrak W}_m(-x_m,\ldots,-x_1)\,\,.
\label{cptt1}
\end{equation}
Under the usual temperedness assumption, the proof of the equality (\ref{cptt1}) as given
by Jost~\cite{Jost} starts of the weak local commutativity (WLC) condition, namely under
the condition that the vacuum expectation value of the commutator of $n$ scalar fields
vanishes outside the light cone, which in terms of Wightman functions takes the form
\begin{equation}
{\mathfrak W}_m(x_1,\ldots,x_m)-{\mathfrak W}_m(x_m,\ldots,x_1)=0\,\,,\quad
{\mbox{for}}\quad x_j-x_{j+1} \in {\mathscr J}_m\,\,.
\label{wlc}
\end{equation}
Jost's proof that the WLC condition (\ref{wlc}) is equivalent to the CPT symmetry
(\ref{cptt1}) one relies on the fact that the proper complex Lorentz group contains
the total spacetime inversion. Therefore, the equality ${\mathfrak W}_n(x_m,\ldots,x_1)=
{\mathfrak W}_n(-x_m,\ldots,-x_1)$ holds, taking in account the symmetry property
${\mathscr J}_m=-{\mathscr J}_m$ in whole extended analyticity domain, by the
Bargman-Hall-Wightman (BHW) theorem. In particular, the BHW theorem has been
shown~\cite{BruNa1} to be applicable to domains of the form ${\mathscr T}_{m-1}=\oR^{4(m-1)}+
V_+(\ell^\prime_{1},\ldots,\ell^\prime_{{m-1}})$. Then, $W^\star_m(\zeta_1,\ldots,\zeta_{m-1})$
can be extended to be a holomorphic function on the extended tube
\[
{\mathscr T}^{\rm ext.}_{m-1}=\Bigl\{(\Lambda\zeta_1,\ldots,\Lambda\zeta_{m-1}))
\mid (\zeta_1,\ldots,\zeta_{m-1}) \in {\mathscr T}_{m-1}, \Lambda \in
{\mathscr L}_+(\oC)\Bigr\}\,\,,
\]
which contains certain real points of type of the Jost points. 

In order to prove that CPT theorem holds in NCQFT, an analogous of the WLC
condition is now formulated:

\begin{definition}
The non-commutative quantum field $\Phi$ defined on the test function space
${\mathfrak H}(T(\oR^4))$ is said to satisfy the weak extended local commutativity (WELC)
condition if the functional
\[
{\boldsymbol{\sf F}}={\mathfrak W}^\star_m(z_1,\ldots,z_m)-
{\mathfrak W}^\star_m(z_n,\ldots,z_1)\,\,,
\]
is carried by set $M_j^{\ell^\prime}=\Bigl\{\bigr(z_1,\ldots,z_m)
\in \oC^{4m} \mid z_j-z_{j+1} \in V^{\ell^\prime} \Bigr\}$.
\label{WELC}
\end{definition}

The WELC condition takes the form $W^\star_m(\zeta_1,\ldots,\zeta_{m-1})-
W^\star_{m}(-\zeta_{m-1},\ldots,-\zeta_1)$ in terms of the NC Wightman functions
depending on the relative coordinates $\zeta_j=z_j-z_{j+1} \in V^{\ell^\prime}$.

\begin{proposition}
Consider $W^\star_m(\zeta_1,\ldots,\zeta_{m-1})$ and
$W^\star_{m}(-\zeta_{m-1},\ldots,-\zeta_1)$. Then 
\[
W^\star_m(\zeta_1,\ldots,\zeta_{m-1})=W^\star_{m}(-\zeta_{m-1},\ldots,-\zeta_1)\,\,,
\]
on their respective domains of holomorphy.
\label{sym}
\end{proposition}

\begin{proof}
The idea of the proof follows from the standard strategy. As in Ref.~\cite{SW}
suppose that $x_1,\ldots,x_m$ are such that all the differences $x_i-x_j$ are space-like.
Then $(z_1,\ldots,z_m) \notin M_j^{\ell^\prime}$. Hence,
\[
W^\star_m(\zeta_1,\ldots,\zeta_{m-1})=
W^\star_{m}(-\zeta_{m-1},\ldots,-\zeta_1)
\]
by Definition \ref{WELC}. Now, our propose is to show that these are points of
holomorphy of both functions. This is achieved applying the Edge of the Wedge
theorem (Theorem \ref{EWTheo}). First, we note that $W^\star_m(\zeta_1,\ldots,\zeta_{m-1})$
is holomorphic in $\oR^{4(m-1)}+V_+(\ell^\prime_{1},\ldots,\ell^\prime_{{m-1}})$
by Corollary \ref{Corol45}. Furthermore,  the functions $W^\star_m(\zeta_1,\ldots,\zeta_{m-1})$
and $W^\star_{m}(-\zeta_{m-1},\ldots,-\zeta_1)$ have boundary values which
agree at totally space-like points in the sense of the strong topology of
${\mathfrak H}^\prime$. Hence, by Theorem~\ref{EWTheo}
$W^\star_{m}(-\zeta_{m-1},\ldots,-\zeta_1)$ is holomorphic at such points. 
\end{proof} 

\begin{theorem}[CPT Theorem]
A non-commutative scalar field theory symmetric under the CPT-operation $\Theta$ is
equivalent to the WELC. 
\label{cpttheo}
\end{theorem}

\begin{proof}
The CPT invariance condition is derived by requiring that the CPT
operator $\Theta$ be antiunitary -- see~\cite{SW}-\cite{Haag}:
\begin{equation}
\langle\Theta\Xi \mid \Theta\Psi\rangle=\langle\Psi \mid \Xi\rangle\,\,.
\label{antiunit}
\end{equation}
This means that the CPT operator leaves invariant all transition probabilities
of the theory. In the case of a NCFT, the operator $\Theta$ can be constructed in
the ordinary way. Taking the vector states as $\langle\Xi \mid\,\,=\langle\Omega_o \mid$ and
$\mid \Psi\rangle=\Phi(z_m)\star \cdots \star \Phi(z_1) \mid \Omega_o\rangle$
we shall express both sides of (\ref{antiunit}) in terms
of NC Wightman functions. For the left-hand side of (\ref{antiunit}) we can use directly
the CPT transformation properties of the field operators, which for a neutral scalar
field is equal to $\Theta \Phi(z)\Theta^{-1}=\Phi(-z)$. Using the
CPT-invariance of the vacuum state, $\Theta \mid \Omega_o\rangle=\,\,\mid \Omega_o\rangle$,
the left-hand side of (\ref{antiunit}) becomes:
\begin{align}
\langle\Theta\Xi \mid \Theta\Psi\rangle &=
\langle\Theta\Omega_o \mid \Theta(\Phi(z_m)\star \cdots \star
\Phi(z_1) \mid \Omega_o\rangle \nonumber\\[3mm]
&={\mathfrak W}^\star_m(-z_m,\ldots,-z_1)\,\,.
\label{thetaw2}
\end{align}

In order to express the right-hand side of (\ref{antiunit}), we take the
Hermitian conjugates of the vectors $\mid \Psi\rangle$ and $\langle\Xi \mid$, to obtain:
\begin{equation}
\langle\Psi \mid \Xi\rangle={\mathfrak W}^\star_m(z_1,\ldots,z_m)\,\,.
\label{thetaw1}
\end{equation}
Putting together (\ref{antiunit}) with (\ref{thetaw2}) and (\ref{thetaw1}), we
obtain the CPT invariance condition in terms of NC Wightman functions as
\begin{equation*}
{\mathfrak W}^\star_m(z_1,\ldots,z_m)=
{\mathfrak W}^\star_m(-z_n,\ldots,-z_1)\,\,,
\end{equation*}
which in terms of the NC Wightman functions depending on the relative coordinates
$\zeta_j$ reads
\begin{equation}
W^\star_m(\zeta_1,\ldots,\zeta_m)=W^\star_m(\zeta_{m-1},\ldots,\zeta_1)\,\,,
\label{cpt-wight}
\end{equation}

Then, without giving more details, it should be clear from the Proposition \ref{sym}
that the arguments of Chap. V of Ref.~\cite{Jost1} apply in our case. Hence,
the CPT theorem continues to hold in UHFNCQFT.
\end{proof}

As it is well-known, the Borchers class of a quantum field is a direct
consequence of the CPT theorem. Thus, we have the

\begin{theorem}[Borchers class of quantum fields for a NCQFT]
Suppose $\Phi$ is a field satisfying the assumptions of Theorem \ref{cpttheo}
and $\Theta$ is the corresponding $CPT$-symmetry operator. Suppose $\psi$ is
another field transforming under the same representation of the {\bf twisted}
Poincar\'e group, with the same domain of definition. Suppose that the functional
$\langle\Omega_o \mid \Phi(z_1)\star\cdots\star
\Phi(z_j)\star {\boldsymbol{\psi}}(z)\star\Phi(z_{j+1})\star\cdots\star\Phi(z_m)
\mid \Omega_o\rangle - \langle\Omega_o \mid \Phi(z_m)\star\cdots\star\Phi(z_{j+1})
\star {\boldsymbol{\psi}}(z)\star\Phi(z_j)\star\cdots\star\Phi(z_1)
\mid \Omega_o\rangle$ is carried by $M_j^{\ell^\prime}=\Bigl\{\bigr(z_1,\ldots,z_{m+1})
\in \oC^{4(m+1)} \mid z_j-z_{j+1} \in V^{\ell^\prime} \Bigr\}$.
Then $\Theta$ implements the $CPT$ symmetry for ${\boldsymbol{\psi}}$ as well and the
fields $\Phi,{\boldsymbol{\psi}}$ satisfy the weak extended local commutativity condition.
\label{Theo5}
\end{theorem}

\begin{proof}
The proof is similar to the proof of Theorem 3.4 of Ref.~\cite{Daniel}.
\end{proof}

\begin{corollary}[Transitivity of the WELC]
The weak relative extended local commutativity property is transitive in the sense
that if each of the fields ${\boldsymbol{\psi}}_1,{\boldsymbol{\psi}}_2$ satisfies
the assumptions of Theorem \ref{Theo5}, then there is a CPT-symmetry operator common
to the fields $\{\Phi,{\boldsymbol{\psi}}_1,{\boldsymbol{\psi}}_2\}$ and by
Theorem \ref{cpttheo}, the weak relative extended local commutativity condition is
satisfied not only for $\{{\boldsymbol{\psi}}_1,{\boldsymbol{\psi}}_2\}$ but also for
$\{\Phi,{\boldsymbol{\psi}}_1,{\boldsymbol{\psi}}_2\}$.
\end{corollary}

\begin{theorem}[Spin-Statistics Theorem]
Suppose that $\Phi$ and its Hermitian conjugate $\Phi^*$ satisfy the WELC with
the ``wrong'' connection of spin and statistics. Then $\Phi(x)\Omega_o=
\Phi^*(x)\Omega_o=0$.  
\end{theorem}

\begin{proof}
The arguments of the standard proof apply~\cite{SW}, since the
properties of Lorentz group representations, existence of Jost points
and the analyticity properties of NC Wightman functions are also available in UHFNCQFT.   
\end{proof}

We complete this section with a of the most important results of the axiomatic approach:
the Reconstruction theorem. Based in our analysis, we have the following

\begin{theorem}[Reconstruction theorem to UHFNCQFT]
Suppose that the hypotheses of Theorem 5.1 in~\cite{BruNa1} hold except
that instead of the sequence $\bigl\{{\mathfrak W}_m\bigr\}_{m \in \oN}$
and of the conditions $(R0)-(R5)$, we have the sequence
$\bigl\{{\mathfrak W}^\star_m\bigr\}_{m \in \oN}$ and the conditions
$\boldsymbol{\sf P^\prime_1}-\boldsymbol{\sf P^\prime_5}$. Then the conclusions
of Theorem 5.1 in~\cite{BruNa1} again hold.
\end{theorem}

\section{Concluding Remarks}
\label{SecFinal}
In the present paper, we extend the Wightman axiomatic approach to NCQFT in terms
of tempered ultrahyperfunctions. An important hint in favor of this approach comes
from the fact that the class of UHFNCQFT allows for
the possibility that the off-mass-shell amplitudes can grow at large energies faster
than any polynomial (such behavior is not possible if fields are assumed to be
tempered only). This is relevant since NCQFT stands as an intermediate framework
between string theory and the usual quantum field theory. Here, we restrict ourselves
to the simplest case, that of a single, scalar, Hermitian field $\Phi(x)$ associated
with spinless particles of mass $m>0$. Some results of the ordinary QFT, the existence
of the symmetry CPT and of the Spin-Statistics connection were proved to hold, if we
replace the local commutativity by an {\em extended local commutativity} in the sense
of Br\"uning-Nagamachi~\cite{BruNa1}. We assume (implicitly) the case of a theory with
space-space non-commutativity ($\theta_{0i}=0$). There is still a number of important
questions to be studied based on the ideas of this paper, such as the existence of the
$S$-matrix, a representation of the Jost-Lehmann-Dyson-type, the Reeh-Schlieder property
and so on. Furthemore, as it was pointed out in~\cite{AGVM}, for gauge theories, in
particular the non-commutative QED (NCQED), the questions associated to the Wightman
axioms and their consequences are more involved due to the UV/IR mixing.
As said at the beginning, the existence of hard infrared singularities in the non-planar
sector of the theory, induced by uncancelled quadratic ultraviolet divergences, can
result in one kind of problem: they can destroy the {\em tempered} nature of the Wightman
functions. This result reinforces the hypothesis that the infrared issue in NCFT must
be dealt with another approach. In this case the ultrahyperfunctional approach to NCQFT
could be an interesting step in order to resolve the problem of the UV/IR mixing in NCFT.
This topic is under investigation.\footnote{We are grateful to the referee for
drawing our attention for the importance of studying the problem of the UV/IR mixing
via ultrahyperfunctional formalism.} We hope to report our conclusions on this issue in
a forthcoming paper.

As a last remark, we note the result obtained in~\cite{Green} where has been showed
that the star commutator of $:\,\phi(x)\star \phi(y)\,:$ and $:\,\phi(y)\star \phi(x)\,:$
does not obey the microcausality even for the case in which $\theta_{0i}=0$. However,
we see that this is not the case here. The condition of extended local commutativity
being defined as a continuity condition of the expectation values of the field commutators
in a topology associated to a complex neighborhood of the light cone, it is not
applied to the tempered fields. Hence, for NCQFT in terms of tempered ultrahyperfunctions
no violation of Einstein's causality is ever involved. 

\section*{Acknowledgments.}
One of us (D.H.T.F.) would like to express his gratitude to Professor O. Piguet
and to the Departament of Physics of the Universidade Federal do Esp\'\i rito
Santo (UFES) for the opportunity of serving as Visiting Professor during
2003-2005, where this work has been initiated. 





\begin{thebibliography}{99}
\bibitem{Comment} The act of attempting to modify the Wigtman axioms by proposing another
space of test functions is quite an old subject~\cite{Wight,Wight1}. Several suggestions
have been made to extend the Wightman axioms for the quantum field theory so as to include
a wider class of fields, see for example~\cite{Jaffe,NaMu,BruNa}.
\end{thebibliography}
\end{document}